\documentclass[twocolumn,showpacs,preprintnumbers,prb]{revtex4-1}
\usepackage{graphicx,bm,amsmath,amssymb}

\def\gz{\ifmmode{Z\hskip -4.8pt Z}
    \else{\hbox{$Z\hskip -4.8pt Z$}}\fi}

\newcommand{\be}{\begin{equation}}
\newcommand{\ee}{\end{equation}}
\newcommand{\bea}{\begin{eqnarray}}
\newcommand{\eea}{\end{eqnarray}}

\begin{document}

\title{Comment on ``Relevance of Cu-3d multiplet structure in models of high Tc cuprates''}
\author{A.~A.~Aligia}
\affiliation{Centro At\'{o}mico Bariloche and Instituto Balseiro, Comisi\'{o}n Nacional de Energ\'{\i}a At\'{o}mica, CONICET, 8400 Bariloche, Argentina}

\begin{abstract}
In a recent work [M.  Jiang,  M.  Moeller,  M.  Berciu,  and  G.  A.  Sawatzky,
Phys. Rev. B \textbf{101}, 035151 (2020)], the authors solved a model with a
Cu impurity in an O-2p band as an approximation to the local electronic structure of a hole doped cuprate. 
One of their conclusions is that the ground-state  has only 
$\sim 50$ \% overlap with a Zhang-Rice singlet (ZRS). 
This claim is based on the definition of the ZRS in a different representation, in which the 
charge fluctuations at the Cu site have been eliminated
by a canonical transformation.
The correct interpretation of the results, based on known low-energy reduction procedures for a multiband model 
including 3d$^8$ 
and 3d$^{10}$ configurations of Cu, indicates that this overlap is near 94 \%.

\end{abstract}

\pacs{71.27.+a, 74.72.-h}
\maketitle



Three decades after their discovery of high-$T_c$ cuprate superconductors, the issue of the appropriate minimal
model that correctly describes the low-energy physics is still debated,
as stated by Jiang \textit{et al}.\cite{jiang} In particular, 
the relevance of the Zhang-Rice singlets (ZRSs) at low energies 
is crucial for the debate.\cite{jiang,zr,eme2,zha,rc,bel,sys,fei}
Recent papers were published with opposite views about the role 
of the Zhang-Rice singlets (ZRSs) in tetragonal CuO.\cite{Adolhps,tcuo}

While the models used to describe the cuprates are periodic and include
usually only one Cu d orbital,  Jiang \textit{et al.} considered an impurity
model with only one Cu site embedded in an O band, but included the full
structure of the 3d Cu shell. This study is relevant, because to explain some Raman and photoemission experiments at excitation energies of order of 1 eV or higher, 
more than one d orbital should be included.\cite{jiang,liu,raman}
They calculate in particular, the ground state for two holes, as 
discussed in more detail below.

The most widely used models used to describe the cuprates can be classified in the following four types in decreasing order of number of degrees of freedom: 

i) the three-band Hubbard model $H_{3b}$,\cite{eme,varma} which contains 
the d orbital of Cu 
of $b_1$ symmetry ($x^2-y^2$) and the p orbitals of O aligned in the direction 
of the nearest Cu atoms. Usually it includes on-site repulsions at Cu ($U_d$) and
O ($U_p$), Cu and O interatomic repulsion $U_{pd}$, in addition to the one-particle 
terms (described in Ref. \onlinecite{jiang}): on-site Cu ($\epsilon_{Cu}$) and
O ($\epsilon_{O}$) energies,
nearest-neighbor Cu-O hopping $t_{pd}$ and O-O hopping $t_{pp}$. The other
models are derived from $H_{3b}$.

ii) the spin-fermion model $H_{sf}$,\cite{eme2,sf} obtained from $H_{3b}$ 
after eliminating the Cu-O hopping term $t_{pd}$ by means of a canonical transformation. Only the $d^9$ configuration of Cu is retained, represented 
by a spin 1/2, which interacts with the fermions of two O bands.\cite{note}
An extension of this model to tetragonal CuO was used in Ref. \onlinecite{Adolhps}.

iii) The one-band Hubbard-like model $H_{1b}$,\cite{fei,sch,lili}
derived from $H_{3b}$ using the cell-perturbation method.\cite{bel,fei}

iv) The generalized $t-J$ model $H_{GtJ}$,\cite{zr,sys} which consists of
holes moving in a background of Cu spins 1/2 with antiferromagnetic exchange $J$,
nearest-neighbor hopping $t$ and additional terms of smaller magnitude.
This model is derived as a low-energy effective one for the others.\cite{bel,fei,sys,tcuo}

In the derivation of the last two models an essential role is played by local ZRSs.

To define properly the ZRSs it is important to realize that, as for any
state or operator, while the \emph{physical meaning} of the states should be
the same for all effective Hamiltonians $H_{\text{eff}}$ used to describe
the cuprates, the\emph{\ form} depends on $H_{\text{eff}}$. In order to give
an example with a familiar case, let us consider the simplest half
filled Hubbard model $H_{\text{Hubb}}=-t\sum\limits_{i\delta }c_{i\sigma
}^{\dagger }c_{i+\delta \sigma }+U\sum\limits_{i}n_{i\uparrow
}n_{i\downarrow }$, where $i+\delta $ denotes the nearest neighbors of site $%
i$ and $n_{i\sigma }=c_{i\sigma }^{\dagger }c_{i\sigma }$. It is well known
that a canonical transformation that eliminates the hopping term to second
order in $t$ leads to the effective Heisenberg Hamiltonian $H_{\text{Heiss}%
}=J\sum\limits_{\langle ij\rangle }\mathbf{S}_{i}\cdot $ $\mathbf{S}_{j}$,
with $J=4t^{2}/U$. This Hamiltonian does not contain double occupied sites.
Of course this does not mean that the original Hamiltonian does not contain
double occupied sites. In fact performing the same canonical transformation
to $n_{i\uparrow }n_{i\downarrow }$ one obtains (see for example Ref. 
\onlinecite{rihm} for the procedure) 

\begin{equation}
(n_{i\uparrow }n_{i\downarrow })_{\text{Hubb}}=(t/U)^{2}\sum\limits_{\delta }%
\left[ \frac{1}{2}-2(\mathbf{S}_{i}\cdot \mathbf{S}_{i+\delta })_{\text{Heiss%
}}\right] .  \label{nnhub}
\end{equation}%
Therefore, as expected, unless all nearest neighbors of site $i$ are
ferromagnetically correlated with site $i$, there is a finite double
occupancy at this site that can be calculated with $H_{\text{Heiss}}$. In
fact, although $H_{\text{Heiss}}$ is a pure spin model, it has been used to
calculate charge-charge and bond-bond 
correlation functions in a modified Hubbard model
with alternating on-site energies.\cite{rihm}.

For $H_{3b}$ the ZRS was defined as the lowest two-hole state of a cell
composed of a d orbital of Cu of $b_{1}$ symmetry at a site  and the
Wannier function $w$ of the O p orbitals at the same site Cu site and with the
same symmetry.\cite{fei}∼∼ $w$  has  92 \% overlap [see Eq. (\ref{ove})] with
the linear combination of 2p orbitals involving only the O atoms nearest to
the Cu site, which we denote as $L_{b_{1}}$ following the notation of
Ref. \onlinecite{jiang}. The rest of the Wannier function extends to more distant
O sites with decreasing amplitude. Therefore the ZRS has basically the same form
as the ground state for two holes found by Jiang \textit{et al. }in their
impurity Hamiltonian for realistic parameters.\cite{jiang} This state [Eq.
(5) of Ref. \onlinecite{jiang}] is 

\begin{equation}
|\psi \rangle =a|b_{1}b_{1}\rangle +b|b_{1}L_{b_{1}}\rangle
+c|b_{1}L_{b_{1}}^{\prime }\rangle +d|d^{10}L^{2}\rangle +...,  \label{psi}
\end{equation}%
where $|b_{1}b_{1}\rangle $ represents the state with two d $b_{1}$ holes, $%
|b_{1}L_{b_{1}}\rangle $ corresponds to the singlet between a d $b_{1}$ hole
and $L_{b_{1}}$, $|b_{1}L_{b_{1}}^{\prime }\rangle $ is the same for more
distant O sites, $|d^{10}L^{2}\rangle $ corresponds to the state with two O
holes and ... denotes states which are not included in $H_{3b}$. It is not
stated how much of the state $|d^{10}L^{2}\rangle $ corresponds to double
occupancy of  $L_{b_{1}}$ (this state, which we denote as 
$|L_{b_{1}}^{2}\rangle $ is the only one included in $H_{3b}$) but from
statements in Ref. \onlinecite{jiang}, I
infer that $|d^{10}L^{2}\rangle \approx |L_{b_{1}}^{2}\rangle $.
The reported coefficients (except for a phase) are 

\begin{equation}
|a|^{2}=0.072\text{, }|b|^{2}=0.549\text{, }|c|^{2}=0.054\text{, }%
|d|^{2}=0.275\text{. }  \label{coef}
\end{equation}%
Adding these numbers one concludes that 95 \% of $|\psi \rangle $ is
consistent with a ZRS and the rest correspond to states not included in  $H_{3b}$.

The representation of the ZRS in $H_{sf}$ corresponds to a singlet between a
Cu spin and a hole occupying $w$ at the same site.\cite{zr}
This was the original definition of the ZRS by Zhang and Rice 
\emph{because they were using $H_{sf}$ as the original model.}\cite{zr} 
Depending on details of the low-energy reduction procedure, the singlet can
also be formed using  $L_{b_{1}}$ instead of the 
O Wannier function.\cite{zr,sys,note2} 
The representation of the ZRS in $H_{1b}$ is  $|b_{1}b_{1}\rangle $ and in $H_{GtJ}$ is a hole. I stress that all representations correspond to the
same physical state.

Based on the weight of $|b_{1}L_{b_{1}}\rangle $ in $|\psi \rangle $,
specifically the value $|b|^{2}=0.549$, Jiang \textit{et al.}.\cite{jiang}
conclude that $|\psi \rangle $ "has only about 50 \% overlap with a ZRS". As
explained above this corresponds to one of the two choices of a ZRS for $%
H_{sf}$  but not for $H_{3b}$ or a model with more degrees of freedom. In
fact, it is easy to realize that a calculation similar to that leading to
Eq. (\ref{nnhub}) can be done using the canonical transformation that maps
the low-energy part of $H_{3b}$ to $H_{sf}$, to show that antiferromagnetic
correlations between a Cu spin and the spin of a nearest-neighbor O atom
implies both, a contribution to the double hole occupancy of 
Cu ($|b_{1}b_{1}\rangle $
state) of the order of $t_{pd}^{2}/(U_{d}-\Delta )$ and to double hole occupancy
of the O atom (part of $|L_{b_{1}}^{2}\rangle $) of the order of 
$t_{pd}^{2}/\Delta $,  for $U_{p}=U_{pd}=0$, where $\Delta =\epsilon
_{O}-\epsilon _{Cu}$.\cite{sf} These contributions are not small, since 
$t_{pd}$ is smaller but of the order of magnitude of the denominators 
$\Delta $ and $U_{d}-\Delta $.\cite{note}

This is of course expected since  $|b_{1}L_{b_{1}}\rangle $ in $H_{sf}$ is
the representation of the the part of $|\psi \rangle $ which contains states
included in $H_{3b}$ and in the first O shell around the Cu impurity. 
It is also natural that experiments in the cuprates identify the 
ZRS as a mixture of doble-hole states $d^9L$, $d^8$, and $d^{10}L^2$, and 
not just $d^9L$.\cite{cuge} 

In the remaining of this work, I made a quantitative comparison of the ZRS
obtained using the cell-perturbation method \cite{fei} and Eqs. (\ref{psi})
and (\ref{coef}). This comparison is in principle expected 
to be only semiquantitative, because  
$H_{3b}$ is a periodic model and in the model of Jiang \textit{et al.} there
is only one Cu atom. Therefore while in $H_{3b}$ the O orbitals tend to form
singlets with their nearest-neighbor Cu atoms, in the impurity model all O
orbitals interact with only one Cu atom. Considering this point, to study the
effect of all the Cu d orbitals in the cuprates it would be probably more
realistic to add them to a localized cell including the full d shell at a Cu
site and the O Wannier function $w$ at that site  than to consider an
impurity model. 

For  $U_{p}=U_{pd}=0$, following Ref. \onlinecite{fei}, the ZRS is obtained as the
ground state of the following matrix involving the states $|b_{1}b_{1}\rangle 
$, the singlet between $b_{1}$ and the Wannier function $|b_{1}w\rangle $
and the doubly occupied Wannier function $|w^{2}\rangle $

\begin{eqnarray}
&&\left( 
\begin{array}{ccc}
U_{d} & V & 0 \\ 
V & \Delta ^{\prime } & V \\ 
0 & V & 2\Delta ^{\prime }
\end{array}
\right) ,  \notag \\
\Delta ^{\prime } &=&\Delta -1.4536t_{pp}, \text{ } 
V =2.7099t_{pd}.  \label{mat}
\end{eqnarray}
Using $U_{d}=8.84$, $\Delta =2.75$, $t_{pd}=1.5$, $t_{pp}=0.55$ 
from Ref. \onlinecite{jiang} and (see Refs. \onlinecite{bel,fei})

\begin{equation}
|\langle w|L_{b_{1}}\rangle |^{2}=0.9180\text{, }|\langle
w|L_{b_{1}}^{\prime }\rangle |^{2}=1-|\langle w|L_{b_{1}}\rangle |^{2},
\label{ove}
\end{equation}%
the ground state of the matrix Eq. (\ref{mat}) can be written in the form 

\begin{eqnarray}
|\text{ZRS}\rangle  &=&a^{\prime }|b_{1}b_{1}\rangle +b^{\prime
}|b_{1}L_{b_{1}}\rangle +c^{\prime }|b_{1}L_{b_{1}}^{\prime }\rangle
+d^{\prime }|L_{b_{1}}^{2}\rangle ,  \notag \\
|a^{\prime }|^{2} &=&0.086\text{, }|b^{\prime }|^{2}=0.583\text{, }%
|c^{\prime }|^{2}=0.052,\notag \\
|d^{\prime }|^{2}&=&0.279.  \label{zrs}
\end{eqnarray}%
Taking into account the difference between the models explained above, the
agreement with Eqs. (\ref{psi}) and (\ref{coef}) is remarkable. 
Assuming $|d^{10}L^{2}\rangle = |L_{b_{1}}^{2}\rangle $, the overlap
is $|\langle \psi |$\text{ZRS}$\rangle |^{2}=0.942$.  In an impurity model, the O
shells that have a hopping with the Wannier function $w$ reduce its
effective energy,  increasing its occupancy and decreasing that of $b_{1}$,
improving the agreement. 

In conclusion, the results of Jiang \textit{et al.},\cite{jiang} 
interpreted correctly on the basis on existing low-energy reduction
procedures from the three-band Hubbard model, support
the validity of generalized Hubbard and $t-J$ models based on 
Zhang-Rice singlets at low energies, even when the 
full Cu-3d multiplet structure (and not just the $x^2-y^2$ orbital) is included.

This work was sponsored by PIP 112-201501-00506 of CONICET, 
and PICT 2017-2726 and PICT 2018-01546 of the ANPCyT, Argentina.


\begin{thebibliography}{99}

\bibitem{jiang} M.  Jiang,  M.  Moeller,  M.  Berciu,  and  G.  A.  Sawatzky,
Relevance of Cu-3d multiplet structure in models of high Tc cuprates,
Phys. Rev. B \textbf{101}, 035151 (2020).

\bibitem{zr} F. C. Zhang and T. M. Rice, 
Effective Hamiltonian for the superconducting Cu oxides,
Phys. Rev. B \textbf{37}, 3759(R) (1988).

\bibitem{eme2} V. J. Emery and G. Reiter, 
Quasiparticles in the copper-oxygen planes of high-Tc superconductors: An exact solution for a ferromagnetic background,
Phys. Rev. B \textbf{38}, 11938(R) (1988).

\bibitem{zha} F. C. Zhang, 
Exact mapping from a two-band model for Cu oxides to the single-band Hubbard model,
Phys. Rev. B \textbf{39}, 7375(R) (1989).

\bibitem{rc} C. D. Batista and A. A. Aligia,
Validity of the $t-J$ model,
Phys. Rev. B \textbf{48}, 4212(R) (1993); \textbf{49}, 6436(E) (1994).

\bibitem{bel} V. I. Belinicher, A. L. Chernyshev, and L. V. Popovich, 
Range of the $t-J$ model parameters for CuO$_2$ planes: 
Experimental data constraints,
Phys. Rev. B \textbf{50}, 13768 (1994), and references therein.

\bibitem{sys} A. A. Aligia, M. E. Simon, and C. D. Batista, 
Systematic derivation of a generalized t-J model
Phys. Rev. B \textbf{49}, 13061 (1994).

\bibitem{fei} L. F. Feiner, J. H. Jefferson, and R. Raimondi, 
Effective single-band models for the high-$T_c$ cuprates. I. Coulomb interactions,
Phys. Rev. B \textbf{53}, 8751 (1996),  references therein.

\bibitem{Adolhps} C. P. J. Adolphs, S. Moser, G. A. Sawatzky, and M. Berciu, 
Non-Zhang-Rice Singlet Character of the First Ionization State of T-CuO,
Phys. Rev. Lett {\bf{116}}, 087002 (2016).

\bibitem{tcuo} I. J. Hamad, L. O. Manuel, and A. A. Aligia, 
Generalized one-band model based on Zhang-Rice singlets for tetragonal CuO,
Phys. Rev. Lett. \textbf{120}, 177001 (2018); references therein.  

\bibitem{liu} R. Liu, D. Salamon, M. V. Klein, S. L. Cooper, W. C. Lee, 
S-W. Cheong, and D. M. Ginsberg,
Novel Raman-active electronic excitations near the charge-transfer gap in 
insulating cuprates,
Phys. Rev. Lett. \textbf{71}, 3709 (1993).

\bibitem{raman} M. E. Simon, A. A. Aligia, C. D. Batista, E. R. Gagliano,
and F. Lema, 
Excitons in insulating cuprates,
Phys. Rev. B \textbf{54}, R3780 (1996).

\bibitem{eme} V. J. Emery, 
Theory of high-$T_c$ superconductivity in oxides,
Phys. Rev. Lett. {\bf{58}}, 2794 (1987).

\bibitem{varma} C. M. Varma, S. Schmitt Rink, and E. Abrahams, 
Charge transfer excitations and superconductivity in “ionic” metals,
Solid State Commun. {\bf{62}}, 681 (1987).


\bibitem{sf} C. Batista and A. A. Aligia, 
Effective Hamiltonian for cuprate superconductors,
Phys. Rev. B \textbf{47}, 8929 (1993).

\bibitem{sch} H.-B. Sch\"uttler and A.J. Fedro, 
Copper-oxygen charge excitations and the effective-single-band theory of cuprate superconductors,
Phys. Rev. B \textbf{45}, 7588(R) (1992).


\bibitem{lili} L. Arrachea and A. A. Aligia,
$d_{x^{2}-y^{2}}$ superconductivity in a generalized Hubbard model, 
Phys. Rev. B 59, 1333 (1999); references therein.


\bibitem{rihm} A. A. Aligia,
Charge dynamics in the Mott insulating phase of the ionic Hubbard model
Phys. Rev. B \textbf{69}, 041101(R) (2004).

\bibitem{note} The parameters obtained from a canonical transformation 
in second order in $t_{pd}$ are not accurate because $t_{pd}$ is not small enough.
Improved parameters of $H_{sf}$ were obtained
fitting the lowest singlet and triplet states of a CuO$_{4}$ cluster
described by $H_{3b}$ and the resulting low-energy model $H_{sf}$ was used
to calculate Cu and O photoemission spectrum of a Cu$_{4}$O$_{8}$ cluster,
obtaining very good agreement with previous results for $H_{3b}$ in the same
cluster.\cite{sf}

\bibitem{note2} Using $L_{b_{1}}$ insted of $w$ to construct the ZRS brings
the technical problem that ZRSs centered at nearest-neighbor Cu sites are not
orthogonal. 


\bibitem{cuge} L.-C. Duda, J. Downes, C. McGuinness, T. Schmitt, A. Augustsson, 
K. E. Smith, G. Dhalenne, and A. Revcolevschi,
Bandlike and excitonic states of oxygen in CuGeO3: 
Observation using polarized resonant soft-x-ray emission spectroscopy,
Phys. Rev. B \textbf{61}, 4186 (2000).

\end{thebibliography}
\end{document}